\PassOptionsToPackage{prologue,dvipsnames}{xcolor}
\documentclass[sigconf,authorversion,nonacm]{acmart}

\AtBeginDocument{%
  }

\copyrightyear{2025}
\acmYear{2025}
\setcopyright{acmlicensed}\acmConference[CHI '25]{CHI Conference on Human Factors in Computing Systems}{April 26-May 1, 2025}{Yokohama, Japan}
\acmBooktitle{CHI Conference on Human Factors in Computing Systems (CHI '25), April 26-May 1, 2025, Yokohama, Japan}
\acmDOI{10.1145/3706598.3713484}
\acmISBN{979-8-4007-1394-1/25/04}

\usepackage{algorithm}  
\usepackage{algpseudocode}
\usepackage{amsmath} 
\usepackage[dvipsnames]{xcolor}
\usepackage{multirow}
\usepackage{listings}
\usepackage{graphicx} 
\usepackage[utf8]{inputenc}

\algnewcommand\algorithmicforeach{\textbf{for each}}
\algdef{S}[FOR]{ForEach}[1]{\algorithmicforeach\ #1\ \algorithmicdo}

\newcommand{\green}[1]{\textcolor{Green}{#1}}
\newcommand{\blue}[1]{\textcolor{CornflowerBlue}{#1}}

\begin{document}

\title{StructVizor: Interactive Profiling of Semi-Structured Textual Data}

\author{Yanwei Huang}
\orcid{0009-0001-9453-7815}
\affiliation{%
  \institution{State Key Lab of CAD\&CG\\Zhejiang University}
  \city{Hangzhou}
  \state{Zhejiang}
  \country{China}
}
\email{huangyw@zju.edu.cn}

\author{Yan Miao}
\orcid{0009-0002-1405-7377}
\affiliation{%
  \institution{State Key Lab of CAD\&CG\\Zhejiang University}
  \city{Hangzhou}
  \state{Zhejiang}
  \country{China}
}
\email{ymiao@zju.edu.cn}

\author{Di Weng}\authornote{Di Weng is the corresponding author.}
\orcid{0000-0003-2712-7274}
\affiliation{%
  \institution{School of Software  Technology\\Zhejiang University}
  \city{Ningbo}
  \state{Zhejiang}
  \country{China}
}
\email{dweng@zju.edu.cn}

\author{Adam Perer}
\orcid{0000-0002-8369-3847}
\affiliation{%
  \institution{Carnegie Mellon University}
  \city{Pittsburgh}
  \state{Pennsylvania}
  \country{USA}
}
\email{adamperer@cmu.edu}

\author{Yingcai Wu}
\orcid{0000-0002-1119-3237}
\affiliation{%
  \institution{State Key Lab of CAD\&CG\\Zhejiang University}
  \city{Hangzhou}
  \state{Zhejiang}
  \country{China}
}
\email{ycwu@zju.edu.cn}

\renewcommand{\shortauthors}{Yanwei Huang, Yan Miao, Di Weng, Adam Perer, and Yingcai Wu}

\begin{abstract}
  Data profiling plays a critical role in understanding the structure of complex datasets and supporting numerous downstream tasks, such as social media analytics and financial fraud detection. While existing research predominantly focuses on structured data formats, a substantial portion of semi-structured textual data still requires ad-hoc and arduous manual profiling to extract and comprehend its internal structures.
  In this work, we propose StructVizor, an interactive profiling system that facilitates sensemaking and transformation of semi-structured textual data. 
  Our tool mainly addresses two challenges: a) extracting and visualizing the diverse structural patterns within data, such as how information is organized or related, and b) enabling users to efficiently perform various wrangling operations on textual data. 
  Through automatic data parsing and structure mining,  StructVizor enables visual analytics of structural patterns, while incorporating novel interactions to enable profile-based data wrangling. 
  A comparative user study involving 12 participants demonstrates the system's usability and its effectiveness in supporting exploratory data analysis and transformation tasks.
\end{abstract}


\begin{teaserfigure}
  \centering
  \includegraphics[width=\linewidth]{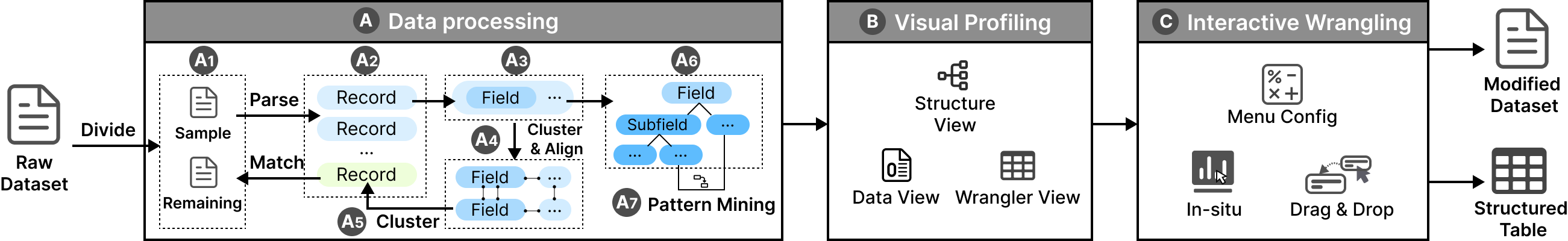}
  \caption{An overview of StructVizor's architecture. (A) StructVizor first divides the raw dataset into a sample set and a remaining set (A1). Using the sample set, it parses data into records (A2), which are then divided into fields (A3). The fields are clustered and aligned (A4) to identify structural patterns of data records, on top of which records can be clustered (A5). The fields can be further iteratively divided into subfields as specified by the user (A6), and relationships between the fields and subfields are then mined to uncover patterns (A7). The profiling results are then visualized on the interface (B), with a variety of interactions provided for users to apply transformations to the dataset or construct relational tables (C). }
  \Description{An overview of StructVizor's architecture. The input is a raw semi-structured textual dataset and the output can be either the modified original dataset or a derived structured table. (A) StructVizor first divides the raw dataset into a sample set and a remaining set (A1). Using the sample set, it parses data into records (A2), which are then divided into fields (A3). The fields are clustered and aligned (A4) to identify structural patterns of data records, on top of which records can be clustered (A5). The fields can be further iteratively divided into subfields as specified by the user (A6), and relationships between the fields and subfields are then mined to uncover patterns (A7). The profiling results are then visualized on the interface (B), with a variety of interactions provided for users to apply transformations to the dataset or construct relational tables (C).}
  \label{fig:overview}
\end{teaserfigure}


\maketitle

\section{Introduction}

Distilling concise metadata such as the representations, patterns, and quality measures from raw datasets, also known as data profiling, is a critical step in the pipelines of data practitioners~\cite{profiling-survey, profiling-revisited, data-profiling}. Data profiles are essential for activities such as assessing data quality, enabling efficient data cleaning, and improving comprehension of data patterns, ensuring that datasets meet analysis requirements~\cite{profiling-revisited}. This foundational step underpins various downstream tasks, including data analysis and model training~\cite{data-profiling, Abdallah2017}.
For instance, in social media analytics, data profiling helps practitioners assess the quality and structural characteristics of user-generated content, enabling the detection of missing values, inconsistent formats, and data distribution anomalies before conducting deeper analytical tasks.
However, despite the existence of many tools or commercial systems for textual data profiling, most of them focus on relational data or standard semi-structured data (e.g., XML, RDF, and JSON)~\cite{profiling-survey, jsoncurer, xml-overview}.
A significant amount of data characterized by heterogeneous, meaningful, and ad-hoc textual structures (e.g., HTML lists and log files), which we refer to as complex semi-structured data, receives considerably less attention.
The diverse structures present in such data poses significant challenges on effective data profiling~\cite{structure-interpretation}.
Accurately detecting and extracting these ad-hoc structures is difficult, and it is equally challenging to visually interpret these structures into actionable insights that facilitate the analysis of textual data.

Prior studies have proposed many automated approaches for detecting structural patterns with heuristic parsing algorithms~\cite{tegra, list-extract, web-record-extract} or program synthesis, including regular expression synthesis, techniques~\cite{flashprog, flashrelate, interpretable-program}. However, these methods typically assume that the given textual data adheres to a uniform and consistent structure, making it difficult to handle heterogeneous data with diverse structural patterns. 
To address this limitation, recent tools like Microsoft SQL Server Data Tools~\cite{MicrosoftSSDT}, Ataccama One~\cite{AtaccamaOne}, and FlashProfile~\cite{flashprofile} divide textual data into several clusters and extract structural patterns as regular expressions for each cluster.
Nevertheless, representing data with one or multiple regular expressions lacks flexibility in controlling pattern extraction granularity (e.g., representing dates as a whole rather than the obscured regular expression \texttt{\textbackslash d\{4\}-\textbackslash d\{2\}-\textbackslash d\{2\}}) and merging semantically-identical patterns (e.g., combining the dates in different formats).
The regular-expression-based approaches also pose challenges for users in interpreting the meaning of complex regular expressions and associating the structural patterns with the corresponding elements in textual data.
Moreover, how the extracted structural patterns can facilitate data understanding and downstream tasks like textual data wrangling remains largely unexplored.


Given the existing limitations, we are motivated to design an interactive interface that enables visual profiling of semi-structured textual data and facilitates textual data wrangling based on data profiles. According to our initial design attempts, two major challenges are identified:
First, \textbf{extracting and visualizing structural patterns for data profiling} can be challenging to due to the diversity of patterns, such as compositions and nested substructures within data records, as well as implicit relationships or dependencies between data segments. Meanwhile, it is equally challenging to \textbf{support various wrangling operations based on data profiles}, as data profiling is often used as a foundational step for data wrangling. Despite prior research on enhancing tabular data wrangling with data profiles~\cite{wrangler}, the potential of profiles on improving wrangling tasks for semi-structured textual data, such as cleaning inconsistent formats and reorganizing components, remains largely unexplored, urging the need for novel interaction designs.

In this paper, we propose StructVizor, an interactive visual profiling system for semi-structured textual data. The system comprises a data processing pipeline that automatically parses input textual data and extracts structural patterns. The resulting data profiles are visualized with an interactive interface, empowering users to understand data structures and uncover actionable insights. Moreover, StructVizor leverages these profiles to facilitate textual data wrangling, enabling users to transform the data format or construct relational tables from specific text segments with novel interactions. 
To evaluate StructVizor, we conducted a user study (N=12) comparing the system to Wrangler~\cite{wrangler}, a representative data wrangling interface with basic profiling functionalities. The results indicated that most participants completed the data wrangling task more quickly with StructVizor. Meanwhile, all participants reported signficantly lower workload except for the mental workload dimension when using StructVizor. Additionally, through an open-ended task on a complicated log dataset, we demonstrated that participants found StructVizor's data profiles and interactions intuitive and effective for explorative data analysis.

The contributions of the paper are as follows.

\begin{itemize}
    \item We propose a novel visual profiling approach for semi-structured textual data combining automatic structure mining and visual pattern interpretation and analysis.

    \item We propose an interactive system, StructVizor, depicting the visual profiling results and supporting profile-based data wrangling, enabling users to perform a diverse range of data transformations in an efficient and seamless way.

    \item We conducted an user study comparing StructVizor with Wrangler to demonstrate the usability and efficiency in data wrangling, while also revealing StructVizor's effectiveness in explorative data analysis through an open-ended task.
\end{itemize}
\section{Related Work}
\subsection{Data profiling}
Defined as the process of acquiring metadata that succinctly characterizes the raw data, data profiling plays a key role in understanding data and significantly influences downstream tasks such as data wrangling and analysis~\cite{data-profiling, profiling-revisited,profiler}. Despite its importance, existing research primarily lies in the context of relational data, a well-structured data format.  Numerous tools have been proposed to handle various profiling tasks, including calculating table statistics, detecting table quality issues, and determining the table schema and patterns~\cite{profiler, potter, cords, trifacta, openrefine, dead-or-alive, datapilot}. For systematic surveys of these systems, see the work by Abedjan et al.~\cite{profiling-survey} and Naumann~\cite{profiling-revisited}. By contrast, fewer research efforts have been devoted to profiling less structured data formats, which are often flexibly and intricately organized and thus much more difficult to handle. Several approaches and systems are available for diagnosing and parsing other standard semi-structured formats like JSON, CSV, XML, and RDF~\cite{pytheas, wrangling-csv, tabular-cell-classification, xml-infer, rdf-profile, jsoncurer}. However, a significant proportion of real-world semi-structured data, such as custom logs and certain HTML lists, remains ad-hoc, exhibiting more random, heterogeneous, and complex structures~\cite{tegra, structure-interpretation, list-extract}.

There have been a few studies on profiling such ad-hoc semi-structured data. Some of these approaches employ heuristic algorithms or program synthesis techniques to generate extraction scripts or regular expressions that describe the data structure~\cite{web-record-extract, tegra, list-extract, flashrelate, flashprog, interpretable-program}. However, these methods assume that the data follows a uniform or predefined structure, which limits their applicability. Recent research studies and commercial products have introduced features that support clustering text and summarizing patterns for each category based on regular expression~\cite{flashprofile, structure-interpretation, semantic-pbe, MicrosoftSSDT, AtaccamaOne}. However, directly presenting the patterns through regular expressions is neither intuitive for comprehension nor effective for detailed analysis, such as comparing subpatterns and associating patterns with the raw data. StructVizor contrasts prior works by offering a visual profiling approach to facilitate structure-related visual analytics. Additionally, it leverages the profiling results as a foundation for downstream tasks such as data wrangling.

\subsection{Visual analytics for textual data}
Understanding large amounts of textual data in a short period is challenging for human beings. Therefore, visual analytics techniques that assist people in understanding and analyzing textual data have been extensively studied~\cite{va-text-survey}. 
However, most existing systems are tailored for general documents, which are often unstructured or with few structural patterns, and there is relatively less focus on structurally organized text. CiteRivers~\cite{citerivers} focuses on bibliographic data and facilitates visual exploration of citation entries. Latif et al.~\cite{vis-author-profiles} present an approach to augment text of authors' profiles with visualizations. Despite their effectiveness, they are designed for text of specific domains. Most relevant to this work, Lee et al.~\cite{tva-audit} propose an interface that allows users to manually process textual data through interactive widgets and provides dashboards to visualize the extracted data entities. However, the manual data preparation process can be demanding for users. StructVizor contrasts these works by enabling automatic data parsing and structure mining, while offering expressive visual profiles to reveal diverse structural patterns.

Additionally, as crucial components in visual analytics systems for textual data, various text visualizations have been proposed to accommodate different analysis requirements, such as keyword searching~\cite{wordle, wordcloud, edwordle, doccards}, topic mining~\cite{textflow, tiara, interpretation-and-trust}, and discourse analysis~\cite{zhao2012facilitating}. However, most text visualizations primarily address the statistical and semantic features of text rather than its syntactic or lexical ones. These latter visualizations rely on specific structural representations and primarily operate at the word level. For example, syntactic patterns are commonly represented using regular expressions, and techniques like glyph collections or statechart diagrams are employed to visualize them~\cite{regex-vis, Regexper}. Strobelt et al.~\cite{lstmvis} annotate words according to the part-of-the-speech tags or vector embeddings generated by machine learning models. Meanwhile, to clearly reveal patterns, visualizations that align these patterns for comparison have been proposed, such as heatmaps, tabular views, and text variant graphs~\cite{dekker2011computer, pockelmann2015catview, silvia2016visualizing}. ~\cite{text-alignment-survey} provides a comprehensive survey on these visualizations.  Built upon these visualizations, StructVizor proposes a visual analytics approach that reveals diverse patterns in semi-structured textual data. It further offers a wider range of interactions to support detailed analysis tasks, such as subpattern inspection and hierarchical views of structures. It also provides interactions for efficient data transformation based on the visual profiling results.

\section{Design goals}
To inform the system design, we have closely collaborated with three experienced data practitioners (P1-P3) whose work heavily involved semi-structured data. P1 was a software testing engineer who frequently analyzed log data to debug or compose performance reports. P2 was a researcher who focused on natural language interfaces for data visualization and was experienced in data wrangling to train language models. P3 was a senior Ph. D. student who researched in semi-structured data profiling for five years. We shared our initial ideas and artifacts generated from the iterative design process with them through bi-weekly meetings over a nine-month period. Additionally, we conducted monthly feedback sessions with a group of twenty students who had varying degrees of experience working with semi-structured data to gather insights and feedback. During these sessions, we presented our design mock-ups or prototypes, and engaged in open-ended discussions to understand their need and gathered suggestions for improvement. Consequently, we have identified the following design goals (DG): 

\textbf{DG1: Automatic, controllable, and transparent data parsing and structural analysis.} 
Initially, our collaborators sought an automatic data processing strategy to eliminate manual parsing and structural analysis. However, through design iterations, we discovered that a fully automatic approach, which attempts to mine all possible patterns without human intervention, would not only lead to an overwhelming amount of calculation but also limit users' ability to inspect and control the data processing pipeline. 
For example, we often received user inquiries regarding the technical details of the processing pipeline. Users expressed a desire for visibility into hyperparameters, such as the granularity of data parsing and clustering, allowing for manual adjustments. Additionally, they wanted the ability to edit profiles to correct any potential errors.

\textbf{DG2: Facilitate overall and detailed sensemaking of common structural patterns through visualizations.} Raw results of structural analysis can be hard to interpret for users. For instance, even experienced users complained to us about the effort spent in interpreting numerous and lengthy regular expressions, urging for intuitive visual aids. Meanwhile, given the abundance of patterns, they also wished to quickly grasp the key information about the structural patterns in the dataset while also having the option to explore interested patterns in detail.


\textbf{DG3: Associate profiling results with raw data.} While data profiling provides a succinct abstraction of the raw dataset, users reported that they could get lost without appropriate explanations of profiling results, such as which parts of the raw data contributed to the given statistics or contained the patterns. Therefore, given the complexity of our system, view coordination ~\cite{nebula} should be carefully designed and cover most system components, faciliating always-on navigation and revealing hidden patterns.

\textbf{DG4: Enable in-situ data wrangling on structures.} In our initial system design, users could extract data segments into a separate view for interactive data wrangling. However, this proved to be inefficient with high interaction numbers. They suggested to integrate a flexible structure view that enables data component manipulation directly through interactions, enhancing in-situ data wrangling. The challenge lies in designing interactions that align with existing views, with each one tailored based on the specific wrangling task for intuitiveness. It is even challenging that the interactions should be expressive enough to support diverse text wrangling operations, whose space, however, remains largely unexplored.

\textbf{DG5: Support table construction from profiling results.} While text transformations can accommodate various scenarios, structured formats like relational tables are still more readily consumed by downstream tasks. Therefore, it is crucial to enable users to construct tables directly from the profiling results.



Additionally, based on DG2 and DG4, we have conducted content surveys of common structural patterns and wrangling operations in semi-structured data management. We discuss the detailed results in Section \ref{strpattern} and \ref{wrangleop}.

\subsection{Structural patterns of semi-structured data} \label{strpattern}
We have conducted a literary review of papers with keywords including semi-structured data, textual data, log data, and data transformation. We also review surveys on relation extraction and text  mining~\cite{nasar2021named,jo2019text}. Since the system takes general semi-structured data as input, we made sure to exclude those focused on domain-specific patterns, such as event sequence patterns in log analysis~\cite{intivisor}, to maintain the focus on the broader aspects of data structures. 
Based on these papers and our experience, we have thus summarized the following structural patterns (SP):

\textbf{SP1: Organization.} One distinguishing characteristic of semi-structured data is its organization as a list of elements, often called \textit{records} in prior literature~\cite{pytheas}, within the data. These records can be further divided into \textit{fields} that carry distinct semantic meanings. The boundaries between records or fields can be either explicit, marked by specific delimiters, or implicit, with no delimiters present~\cite{structure-interpretation}. Additionally, some datasets may also include contextual information, such as headers and footnotes, that does not belong to the metadata~\cite{pytheas}.

\textbf{SP2: Sub-structures.} In addition to the overall structure of the dataset, its components can also exhibit interesting sub-structures. For example, within the fields of log data, there may be special formats such as timestamps, dates, URLs, or even nested structures like JSON objects~\cite{structure-interpretation,wrangling-csv,pytheas,multi-hypothesis-csv}. These sub-structures provide further insights into the organization and content of the data, and understanding them is crucial for effective analysis and interpretation.

\textbf{SP3: Data relationships.} Similar to function dependency in relational databases, relationships are also prevalent in semi-structured data~\cite{logrep, arimura2002efficient, extract-www}. For instance, given a record ``$userId: 95001, path: api/query/95001$'', the path is determined by the corresponding user ID. More complex patterns include values with matched prefixes/suffixes, sequential order, or specific positional spacing~\cite{extract-www, arasu2003extracting, effective-pattern-discovery}. 

\textbf{SP4: Hierarchies.} Several works on textual data mining have stressed the ``granularity'' or ``specificity'' of structural patterns~\cite{flashprofile, extract-www}, showing that data fields are often naturally hierarchical. For example, a date format can be interpreted either as a whole or decomposed into their constituent parts, such as year, month, and day. Even a seemingly simple numerical field can be further categorized based on its digits, depending on the specific scenario. Recognizing and understanding these hierarchical structures can allow for a more nuanced analysis and interpretation of the data.




\begin{figure*}{
  \centering
  \includegraphics[width=\linewidth]{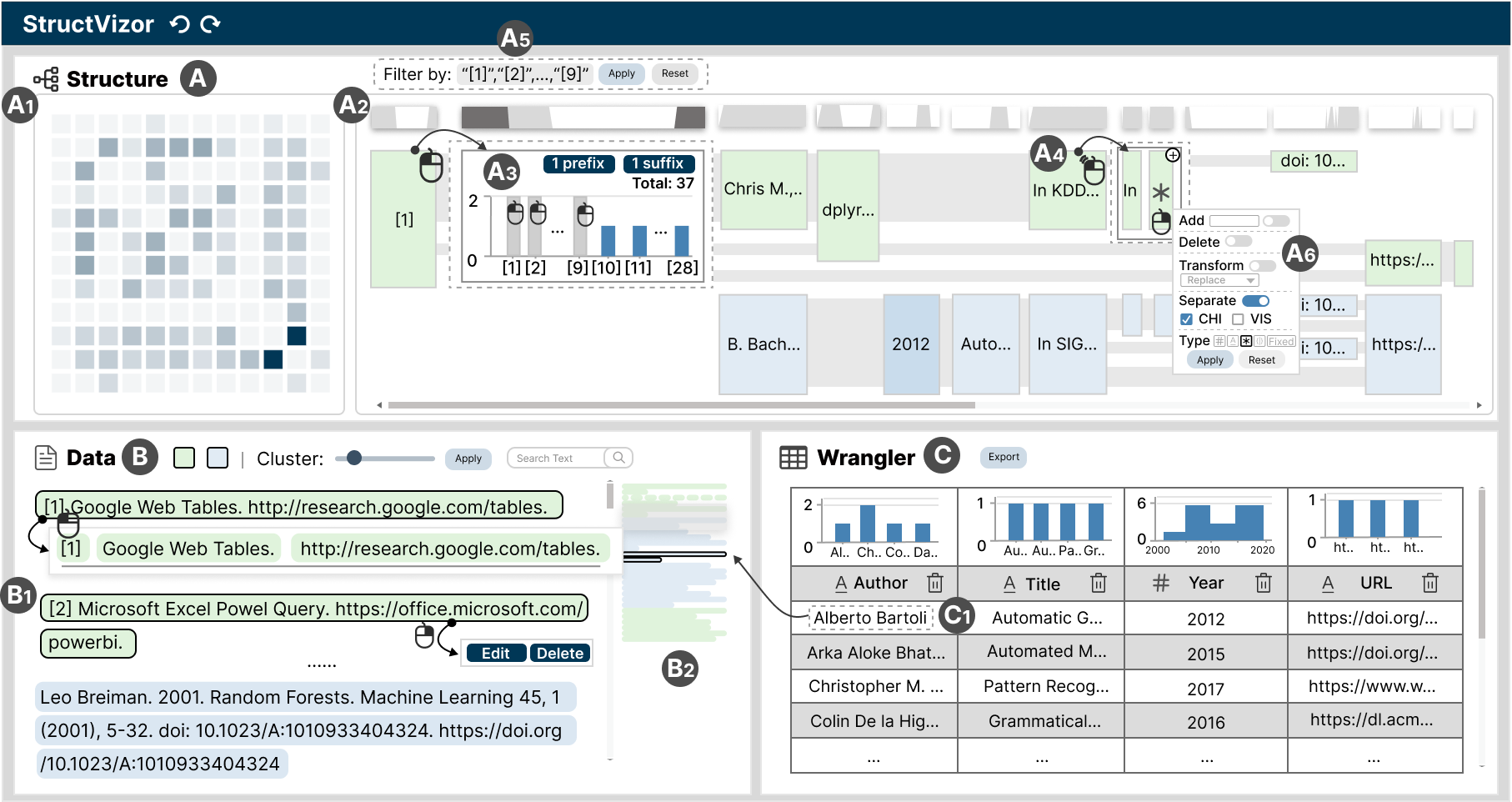}
  \caption{%
  	The StructVizor system. (A) The structure view visualizes the structural patterns present in the dataset. The relationship view (A1) illustrates the similarity between different data fields. The tabular view (A2) depicts the dataset's structural distribution, where rows represent data records and columns represent data fields. Users can click on cells (A3) to view the value distributions of records in the selected field. Various interactions are supported for in-situ data wrangling, such as splitting fields into subfields (A4), applying filters (A5), and performing transformations on cells (A6). (B) The data view displays the annotated dataset, with parsed data records separated into different lines and clustered (B1). An overview of the dataset is provided through the thumbnail view (B2). (C) The Wrangler view empowers users to construct relational tables based on the dataset profiles. Users can navigate to specific records by clicking on the table cells (C1).
  }
  \Description{The StructVizor system. The scenes of each view are the same as the usage scenario.}
  \label{fig:system}
}
\end{figure*}

\subsection{Wrangling operations of semi-structured data} \label{wrangleop}
To cover common wrangling operations for semi-structured data, we compile a set of operations by drawing from existing literature on data wrangling for tabular data~\cite{wrangler, rigel, nl2rigel, table-scraps} and standard semi-structured data formats~\cite{jsoncurer, hitailor}. We also incorporate common text transformation operations found in libraries such as Pandas~\cite{pandas}. 






\begin{itemize}

    \item \textit{Create.} Add content to the dataset, such as including new data records or enclosing words with additional quotation marks.

    \item \textit{Delete.} Remove content from the dataset, which may involve deleting leading spaces, filtering subsets of data, or organizing extracted data into a new structured file.

    \item \textit{Transform.} Convert data records or fields in a one-to-one manner. This includes string replacement, capitalization, and pattern testing, often using regular expressions.

    \item \textit{Separate.} Map selected items to multiple values, typically using functions like split to divide strings into components.

    \item \textit{Combine.} Merge multiple records or fields into a single value. This encompasses operations such as concatenation, aggregation, and joining to consolidate data from various sources.

    \item \textit{Schema.} Adjust the dataset based on its abstracted schema rather than data values. This can include removing redundant keys in JSON~\cite{jsoncurer}, rearranging field orders, or changing the identified data types for each field.
\end{itemize}


\begin{figure*}[t]
  \centering
  \includegraphics[width=\linewidth]{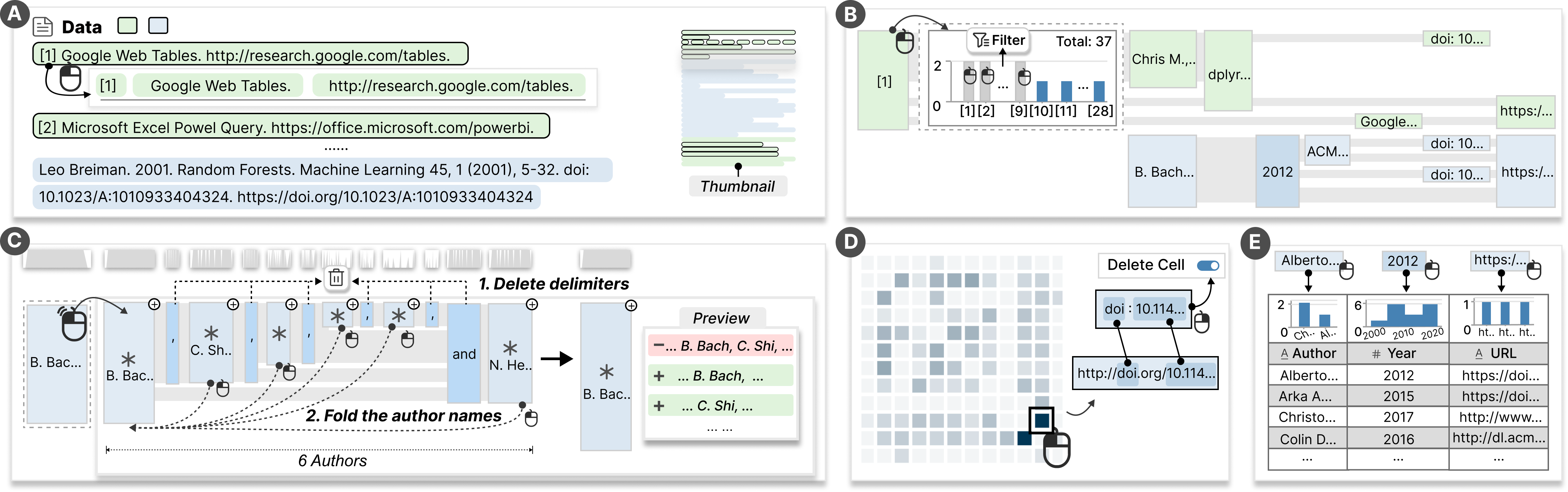}
  \caption{Scenes for the usage scenario. (A) The data view after importing the dataset. (B) Cells in the structure view are used for field analysis and data filtering. (C) The updated structure view for detailed analysis of lengthy unstructured strings. (D) The heatmap showing the similarity between fields. (E) The relational table  in the wrangler view constructed by the user in for further analysis.}
  \Description{(A) The data view contains green and blue records. Users can click on the green record (e.g., ``[1] Google Web Tables. http://research.google.com/tables.'') and to see the decompose fields (e.g., ``[1]'', ``Google Web Tables.'', and ``http://research.google.com/tables.''). (B) The user clicks on the citation index cell, and a bar chart is displayed, where the first nine indices appear twice while others appear once. (C) The tabular view showing the comma-separated substructure of author lists. Users perform several interactive transformations on it, with a preview window showing the results. (D) The heatmap view showing the similarity between fields DOI and URL. (E) The user drags three fields to the wrangler view to construct a table, namely author, year, and URL. }
  \label{fig:scenario}
\end{figure*}

\section{Usage Scenario}

Informed by the design goals, we designed StructVizor, a visual profiling system of complex semi-structured data. \autoref{fig:system} illustrates the interface of StructVizor, which consists of three views: a data view (B) displaying the raw dataset, a structure view (A) showing mined structural patterns, and a wrangler view (C) for table construction after analysis. 

Next, we walk through its usage through a scenario, where a data analyst, Peter, is going to use StructVizor to analyze and clean a citation dataset with multiple data issues. The dataset contains citation entries of heterogeneous structures, covering various information such as authors, publication years, DOIs, and so on. Note that the dataset is in plain text, formatted as a concatenation of citation entries.

Peter begins by importing the dataset into StructVizor, which automatically parses the dataset, mines the data patterns, and visualizes the results on the interface. He initially examines the data records in the data view (\autoref{fig:scenario} (A)) and observes that the records are grouped into two clusters, indicated by the colors \green{green} and \blue{blue}. From the thumbnail view on the right, Peter observes that the  \green{green} records primarily appear at the beginning and end of the dataset, while the central part contains \blue{blue} records. He clicks on the first \green{green} record and finds that it contains three fields: a citation index, a title, and a URL. Similarly, he selects an arbitrary \blue{blue} record and discovers some additional fields within it, such as authors, venues, and DOI. This exploration gives him a preliminary understanding that records in different clusters have significantly distinct structures.

Next, Peter switches to the structure view (\autoref{fig:scenario} (B)) to investigate the detailed structures. In this view, each row represents a record, each column represents a field, and each cell represents the value of a specific field in the corresponding record. Cell heights indicate the number of records containing this field, and the color encoding is the same with the data view. He discovers that all \green{green} records commence with a citation index (e.g., \textit{[1]}), and most records include a list of authors (e.g., \textit{Chris. M.,...}) and a title (e.g., \textit{dplyr...}). These records also have an optional DOI field (e.g., \textit{doi: 10...}) or a URL (e.g., \textit{https:/...}). Meanwhile, by following the data flow, he further notices a few \green{green} records that contain a  corporation name (e.g., \textit{Google...}) after the citation index, followed by a URL. By contrast, all \blue{blue} records start with an author list (e.g., \textit{B. Bach...}) and a year (e.g., \textit{2012}) while ending with a URL. In addition, a few \blue{blue} records contain a publisher (e.g., \textit{ACM}) and a DOI. The structure view clearly shows that records of different colors have significantly different field compositions, reinforcing his earlier hypothesis.

With an overview of the dataset in mind, Peter decides to dive into each field for exploration. He starts by clicking on the first field, and a bar chart is generated showing the distribution of the citation indices. He finds that indices \textit{[1]} to \textit{[9]} appear twice, more frequently than the other ones. To figure out the reason, he clicks on the bars for these indices in the chart to indicate a filter, and all \green{green} records starting with the selected indices are highlighted in the data view (\autoref{fig:scenario} (A)). Upon inspecting the dataset, he discovers that the first nine records are duplicated at the end of the dataset. Consequently, he removes these redundant records.

Next, Peter moves on to inspect the author lists. By double-clicking on the cell, StructVizor reveals the detailed composition of this field (\autoref{fig:scenario} (C)). He discovers that the author list comprises multiple arbitrary strings (where \includegraphics[width=0.2cm]{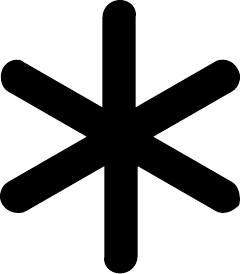} represents arbitrary characters and \includegraphics[width=0.2cm]{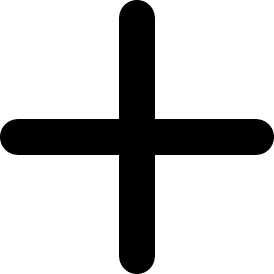} indicates characters appearing multiple times) separated by commas or the word ``\textit{and}''. The number of these strings ranges from one to six, suggesting that the citations include between one and six authors. Due to the variability and complexity of the author list's substructure, Peter first deletes the delimiters (\autoref{fig:scenario} (C1)) and then drags individual cells together to merge them (\autoref{fig:scenario} (C2)). As shown in the consequent preview panel, StructVizor will generate a new record for each author while keeping the other parts of the record unchanged.

After reviewing several fields for insights, Peter shifts his attention to the heatmap view (\autoref{fig:scenario} (D)). He notices a few grids with dark colors, indicating strong relationships between pairs of fields. When he hovers over one of these grids, StructVizor highlights the corresponding fields in the tabular view. He discovers that the URL field includes the DOI content string as a substring, which explains the strong dependency. To eliminate this redundancy, Peter right-clicks on the DOI field and selects ``\textit{Delete Cell}''. Finally, he drags several cells that he examined from the tabular view  into the wrangler view to create a relational table (\autoref{fig:scenario} (E)). He then exports the table for his analysis report.

\section{StructVizor}

This section introduces the detailed design of StructVizor. \autoref{fig:overview} shows the overall architecture of StructVizor. Taking a semi-structured dataset as input, the system first processes data for structure parsing and pattern mining (A). It then visualizes the profiling results on an interface (B) where users can interactively inspect the profiles or perform wrangling operations (C) to modify the raw dataset or construct a new table.


\subsection{Data processing}
Following DG1, StructVizor is built upon a data processing module that automatically parses the raw dataset and analyzes the structural patterns. DG1 also requires that data with heterogeneous organizations or diverse structural patterns should be supported. However, most prior works on structural interpretation of semi-structured data assume that the data follows a single or uniform organization style~\cite{structure-interpretation, pytheas}. We have also noticed two exceptions: UnRavel~\cite{flashprofile} and DataMaran~\cite{datamaran}, which can potentially support data with multiple structural templates. However, both rely heavily on structural assumptions (e.g., records and fields being delimited by explicit symbols), and neither tool is open-sourced. This motivates us to design a data processing pipeline that operates without specific assumptions.

Data processing starts by dividing the raw dataset into a sample set and a remaining set (\autoref{fig:overview} (A1)). The sample set is initially used for parsing and pattern mining, while similar steps are subsequently performed on the remaining set with the assistance of the previous results. This ensures the scalability of the approach as the following data parsing involves large language models which can only accept a small input size. In data parsing, the input data is decomposed into a series of \textit{records} (\autoref{fig:overview} (A2)). Each record can be divided into \textit{fields} (\autoref{fig:overview} (A3)), which can be further divided into subfields when specified by the user (\autoref{fig:overview} (A6)).
During structural pattern analysis, the system syntactically clusters the fields in different records for alignment (\autoref{fig:overview} (A4)), and uses the alignment results to cluster the parsed records (\autoref{fig:overview} (A5)). It further mines patterns including regular expresssions and data dependency on the field and subfield levels (\autoref{fig:overview} (A7)). Next, we discuss the details of the data processing pipeline.

\subsubsection{Data sampling} 
\begin{figure*}[t]
  \centering
  \includegraphics[width=0.95\linewidth]{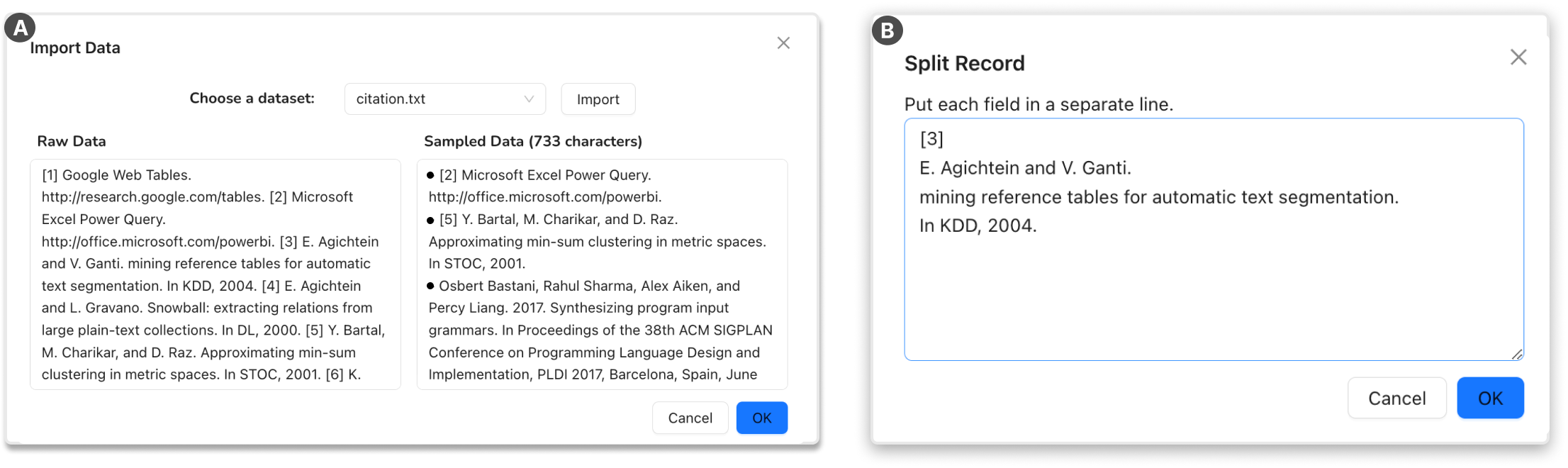}
  \caption{(A) The panel for importing data. The automatically sampled dataset is initially shown. Users may edit and refine it before entering the visualization interface. (B) The panel for editing the fields of a record, where users can put the updated fields in separate lines. }
  \Description{(A) Left: the raw textual dataset. Right: the sampled dataset. (B) A text panel for editing the fields of a record, where users can put the updated fields in separate lines. }
  \label{fig:import}
\end{figure*}

Given the diverse structures within data, generating a sample that consists of all major structural patterns is non-trivial. We propose a sampling approach that begins with randomly sampling several substrings from different parts of the dataset. We make sure that the maximum total length of these substrings is 4,000 characters to accommodate large datasets. Since these substrings may contain broken records, we prompt the GPT-4o model to capture all complete records within them, taking advantage of its ability to understand subtle semantic boundaries between records. The identified records will form the final sample, which users can further edit and refine through an interactive panel (\autoref{fig:import} (A)) beforing entering the visual profiling interface (DG1). The prompt template we use for sampling is: \texttt{Textual data can be divided into a list of records, which are substrings and are semantic units. The following data may include incomplete records. Parse the given data and output all complete records you could find, return only a JSON array where all elements are enclosed in double quotes. Do not add any content before or after this JSON array. Data: \textless sampled\_string\textgreater }.

\subsubsection{Data parsing}
Data records are segments of the raw dataset with similar structures. For instance, each row in a standard CSV file can be viewed as a data record. While some semi-structured data has explicit delimiters for data records (e.g., \textit{enter} in HTML lists), the situation becomes more complex for ad-hoc formats. One typical example is \textit{implicit delimiters}~\cite{tegra}: assume a space-separated file with two records ``\textit{01 Olivia Mitchell Female 02 Ethan Anderson Male}''. In this case, spaces serve not only as separators of records but also as separators of words, making it infeasible to directly separate records and fields using spaces alone. Additionally, in some extreme cases, there may be multiple delimiters or even no delimiter at all.

To take the semantical context into consideration, we leverage GPT-4o to divide the input data into records and fields. The prompt template we use is: \texttt{Textual data can be divided into a list of records, where each record can be further divided into several fields. Fields are substrings of a record and are semantic units. Delimiters (e.g., enter) should be preserved as individual fields so that a record can be retained by concatenating all fields. Parse the following data into records and fields, and output them as a JSON (format: [\{``fields'': [``...'', ...], ..., ...], where each object represents a record). Data: \textless dataset\textgreater }. Similarly, fields can then be iteratively divided into subfields with the following prompt template: \texttt{Fields can be divided further into a list of subfields, which are semantic substrings of inputs. Split each field into subfields. Delimiters (e.g., enter) should be preserved as individual subfields so that a field can be retained by concatenating all subfields. In the following data, each line represents a field. Parse these fields into subfields, and output them as a JSON (format: [{``subfields'': [``...'', ...]}, ...], where each object represents a field). Data: \textless dataset\textgreater}. Note that the subfields are calculated only when specified by the user due to the potentially large number of fields.


\subsubsection{Structural pattern analysis}
The process of structural pattern analysis comprises of three steps:  field alignment, record clustering, and pattern mining.

\textbf{Field alignment.} Informed by DG2, StructVizor aligns fields with similar structural patterns for profiling. For simplicity we mainly consider the syntactical structure of fields during the alignment and leave other structural patterns for future work. Specifically, we leverage Microsoft PROSE SDK~\cite{PROSE} to cluster the values of all fields and extract regular expressions for each cluster. During the alignment, only fields within the same cluster can be aligned. The alignment process is essentially a \textit{multiple sequence alignment} (MSA) problem~\cite{msa}. However, MSA is known to be NP-complete with the computational time of an optimal solution growing exponentially with the record number~\cite{msa-proof}. To balance between efficiency and performance, we follow a greedy approach where we iterate over all records and align the $n$-th record with the previously aligned $n-1$ records, which can be easily solved using dynamic programming. The pseudo-code for this approach is outlined in Algorithm \ref{pseudo-code}.

\begin{algorithm}[t]
\caption{The field alignment algorithm}
\label{pseudo-code}
\begin{algorithmic}[1]
\Statex \textbf{Input} \hspace{0.5em} \textbf{:} \text{an array of records $R$}
\Statex \textbf{Output} \textbf{:} \text{a 2-dimensional array $T$ indicating the aligned table}
\Procedure{alignFields}{$R$}
\State $N \gets \Call{len}{R}$
\State $T \gets \text{Array}[N][] $ \Comment{Dynamic length}
\State $T[0] \gets R[0] $
\For{$i=1$ to $N-1$}
    \State $indexesOfAlignedColumns \gets$ \Call{align}{$R[i]$, $R[0..i-1]$} \Comment{The optimal alignment  by dynamic programming}
    \For{$j=0$ to \Call{len}{$R[i]$}}
    \If {$j \in indexesOfAlignedColumns$}
    
    \quad\quad\quad \Call{updateTable}{$T$, $newColumn=false$}
    \Comment{Put $R[i].fields[j]$ and the aligned fields in the same column}
    \Else
    
    \quad\quad\quad \Call{updateTable}{$T$, $newColumn=true$}
    \Comment{Create a new blank column and insert $R[i].fields[j]$ into it}
    \EndIf
    \EndFor
\EndFor

\State \Return $T$
\EndProcedure
\end{algorithmic}
\end{algorithm}


\textbf{Record clustering.} Similarly following DG2, StructVizor clusters the records to help users efficiently distinguish between differently structured records in the data overview. Based on the field alignment results, records can be viewed as aligned sequences of equal lengths and their differences can be measured with Hamming Distance. Formally, let $M$ be the number of columns derived from the alignment, the distance between any pair of records $(r_i, r_j)$ is defined as:

\[
dist(r_i, r_j) = \sum_{k=1}^{M} r_{ik} \oplus r_{jk}
\]

where the value of $a \oplus b$ is 0 if $a=b$ and 1 otherwise. We leverage the DBSCAN algorithm to cluster the records, with hyperparameters like $eps$ customizable by users.

\textbf{Pattern mining.} Informed by DG1 and DG2, StructVizor conducts pattern mining on the data to facilitate users' sensemaking. While previous steps have revealed the organization manner of data and the hierarchical representation of fields, this step aims to mine additional common patterns to cover the space outlined in Section \ref{strpattern}. Specifically, fields or subfields of common special formats such as URL, DOI, and ISBN are extracted through their regular expression-based representations learned in the clustering process. We also mine string prefix and suffix patterns for field values through enumeration. In addition, various mining approaches can be used to determine different relationships between fields or subfields. Currently, we focus on string similarity, but this approach can be easily extended to encompass more intricate patterns by replacing the scoring function below with relevant metrics. Formally, given the record set $R$, the similarity score of two fields $a$ and $b$, or $S(a, b)$, can be calculated by:

\[ S(a, b) = \frac{\alpha}{|R'|} \sum_{r \in R'} (1- \frac{L(r[a], r[b])}{max(|r[a]|, |r[b]|)}) \]

where $L$ represents the Levenshtein distance, and $r[a]$ and $r[b]$ represent the strings of fields $a$ and $b$ in record $r$. $R'$ is the set of records that contain both the fields $a$ and $b$:

\[ R' = \{r\in R\ |\ r[a]\neq \emptyset, r[b]\neq \emptyset\} \]

Furthermore, an adjusting factor, denoted as $\alpha = log_{10}(|R'|)$, is introduced to scale the similarity score so that the score will not become excessively large when only a few unique value pairs exist. 

\subsubsection{Processing the remaining dataset} Despite the previous steps for processing the sample dataset, promoting the results to the remaining dataset is still non-trivial. Most existing approaches handle large datasets by synthesizing programs from the sample data and running them on the remaining data. However, when dealing with an ad-hoc and heterogeneous dataset, finding a perfect sample can be challenging, as new structural patterns are inevitable, leading to potential failures of the synthesized programs. Furthermore, the organization of the remaining dataset, such as the boundaries between records and fields, remains unknown, making the parsing process highly challenging. 

In view of this, we propose a more robust and adaptive approach that models the processing of the remaining data as an optimization problem. Drawing inspiration from DataMaran~\cite{datamaran}, we adopt a template matching approach: each record in the processed sample dataset can be viewed as a raw structural template. The templates can be pruned, extended, or modified to match the records in the remaining set. Our goal is to maximize the number of matched fields while minimizing the number of characters that do not belong to any record. Formally, let the number of matched fields be $MF$ and the number of unmatched characters be $UC$, we define the objective function $f$ as:

\[ f = MF - \alpha \cdot UC \]

\noindent where $\alpha = 0.01$ by default. One can prove that finding the optimal $f$ is NP-hard (see the appendix for a proof), and we realized through our initial attempts that searching for a global optimal solution can be time-consuming. To balance between performance and time, we propose a novel algorithm which iteratively finds the locally optimal template for matching, as outlined in Algorithm \ref{parse-remaining-data}.  The local matching problem can be solved through dynamic programming in a straightforward way. Note that when matching the data with templates, we leverage \textit{fuzzy matching} through the Regex library~\cite{Regex}, allowing variations of the raw templates to accommodate the heterogeneous data. A match is considered valid only when the minimum number of changes required to the template for matching is within a predefined threshold (3 changes for each field in our implementation). In addition, we have proposed several optimizations to further expedite the algorithm, and we refer readers to the appendix for more details.

\begin{algorithm}[t]
\caption{Algorithm for parsing the remaining dataset}
\label{parse-remaining-data}
\begin{algorithmic}[1]
\Statex \textbf{Input} \hspace{0.5em} \textbf{:} \text{a list of structural templates $ST$ and the remaining} \\
\text{dataset $D$}
\Statex \textbf{Output} \textbf{:} \text{a 2-dimensional array $R$ indicating the parsed data} \\ \text{records and fields}
\Procedure{parseRemainingData}{$ST, D$}
\State $R \gets []$
\State $N \gets$ \Call{len}{$D$}
\State $currentIndex \gets 0$
\While{$currentIndex < N$}
\State $record, nextIndex \gets$ \Call{findLocallyOptimalMatching}{$ST, D[currentIndex..N-1]$}
\State $R$.\Call{Append}{$record$}
\State $currentIndex \gets nextIndex$
\EndWhile

\State \Return $R$
\EndProcedure
\end{algorithmic}
\end{algorithm}

\subsection{Visual profiling}

After the data parsing and pattern mining, the profiles are visualized on the interface of StructVizor (DG2). \autoref{fig:system} illustrates the interface of StructVizor, which consists of three views: a data view (B), a structure view (A), and a wrangler view (C).

\subsubsection{Data view}
The data view is designed to provide an annotated view of the raw dataset. The parsed data records are wrapped in different colors based on the clustering results (\autoref{fig:system} (B1)). Since the dataset can be very large, a thumbnail view is provided as an overview (\autoref{fig:system} (B2)) where users can drag the scrollbar on it to navigate within the data. While only the records are displayed by default, users can click on the records to inspect the detailed fields. Similarly, users can click on the fields to view the subfields. Moreover, a slider is available to adjust the granularity of record clustering.

In addition, the data view is equipped with several interactions to support profile modification in case of potential errors (DG1). Users can right-click on a data record and select the \textit{Edit} option from the menu to adjust parsing results in the result panel (\autoref{fig:import} (B)). This action opens a window where each field appears in a separate row within an editable text area. Users also have the option to delete a record or a cluster via the menu. Note that after an edit, only the modified record or field will be re-rendered with other parts of the system unchanged. Whenever the user makes an edit, an ``\textit{Apply changes}'' button will appear at the header of the data view. Users can click on it after finishing all desired local edits, and StructVizor will update other views by first adding the updated records to the sample set, and then rerunning the entire data processing pipeline except data sampling and data parsing.

\subsubsection{Structure view}
Designed to visualize the structural patterns, the structure view includes a relationship view (\autoref{fig:system} (A1)) and a tabular view (\autoref{fig:system} (A2)). 


\textbf{Relationship view.} The relationship view incorporates a heatmap visualizing the connections between fields. Each row and column represents a field and the opacity of each cell encodes the similarity between a pair of fields. The heatmap supports zooming in and out, which is especially useful when there are numerous fields to analyze.  When users hover over the cells with their mouse, the corresponding fields will be highlighted in the tabular view. 

\textbf{Tabular view.} The tabular view visualizes the table acquired in the field alignment process. Each row represents a record, each column represents a field, and each cell represents the value of a specific field in the corresponding record. The table is sorted sequentially, starting from the first field and progressing towards the last field. To enhance the readability and reduce visual clutter, adjacent cells whose corresponding records belong to the same cluster are merged into a single cell. The merged cell displays the value from the field of the first record in the cluster. The color of a cell represents the cluster to which the corresponding record belongs, while the opacity of the cell indicates the percentage of unique values within the merged values. Cells that share common records are visually connected to facilitate easier navigation.  Additionally, since the original data order is not preserved in the tabular view, each column in the tabular view is accompanied by an area chart above it. The area chart represents the positional distribution of the selected field's values in the original dataset. For instance, the area charts in \autoref{fig:system} (A2) indicate that records containing the fields in the first two columns are  located at the beginning and the end of the dataset. This  representation helps  associate the columns with the original dataset and provides general distribution information. 

Moreover, the tabular view is equipped with multiple interactions to facilitate explorative data analysis.
Users can click on a cell to view the value distribution, with a bar chart for categorical fields or a histogram for quantitative fields (\autoref{fig:system} (A3)). The cell also contains two buttons where users can hover on to view the string prefix and suffix patterns within the cell values. Meanwhile, users can double click on a field cell to expand and view the subfields (\autoref{fig:system} (A4)). The subfield cells are embedded with icons or text that correspond to each component of the field's regular expression representation. Specifically,
{\includegraphics[width=0.2cm]{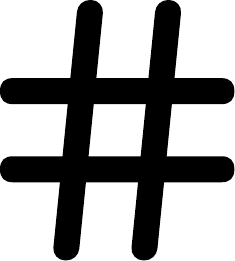}} represents a digit, {\includegraphics[width=0.25cm]{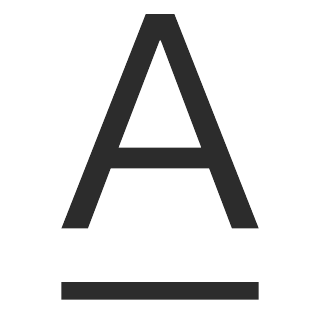}} represents a letter, {\includegraphics[width=0.2cm]{figures/inline/any.pdf}} represents an arbitrary character, {\includegraphics[width=0.25cm]{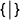}} denotes one of several optional elements, and plain text refers to fixed strings. Textual annotations are also provided for special formats like CSV, JSON, and URL. The superscript at the top right of the cell indicates the number of repeats for the element in the cell, which can be either a numerical value or a {\includegraphics[width=0.2cm]{figures/inline/plus.pdf}} icon to indicate that the element will be repeated an arbitrary number of times. 

Note that one of the major drawbacks of using tabular layouts for text alignment is the lack of scalability~\cite{text-alignment-survey}, as the height of individual cells can become excessively large when dealing with large datasets. To address this issue, we not only support zoom-in/out interactions to scale the view but also propose a novel layout algorithm whose main idea is dividing the merged cells into smaller units based on their positions. Each of these smaller cells is assigned a height proportional to the logarithmic value of the number of cells merged. We also set a minimum cell height to prevent small cells from becoming invisible. The height of the cell is determined by the combined height of its divided cells. This method ensures that the overall cell height grows approximately logarithmically with the data size to accommodate the limited screen space while also preserving the table structure. Algorithm ~\autoref{layout} shows the pseudo-code. The algorithm first iterates over the cells and records the starting and ending row indices for each cell (lines 2-6). In a subsequent iteration, it divides each cell into subcells using these indices (lines 7-8). Each subcell is assigned a height logarithmically proportional to its size, and the overall cell height is calculated by aggregating all subcell heights (lines 9-13).

\begin{algorithm}[t]
\caption{The layout algorithm for the tabular view}
\label{layout}
\begin{algorithmic}[1]
\Statex \textbf{Input} \hspace{0.5em} \textbf{:} \text{an array of the cells in the tabular view $cells$}
\Statex \textbf{Output} \textbf{:} \text{the updated cells $cells$ with calculated height $height$} \\ \text{for each cell.}
\Procedure{ParseData}{$cells$}
\State $rowIndexForDivision\gets$ $\{ \}$
\ForEach{$ cell \in cells$}
\State \Call{insert}{$rowIndexForDivision$, $cell.firstRowIndex$}
\State \Call{insert}{$rowIndexForDivision$, $cell.lastRowIndex$}
\EndFor
\ForEach{$ cell \in cells$}
\State $subCells\gets$  \Call{divide}{$cell$, $rowIndexForDivision$}
\State $cellHeight\gets$  \text{$0$}
\ForEach{$ subCell \in subCells$}
\State $subCellSize\gets$ $subCell.lastRowIndex - subCell.firstRowIndex$
\State $cellHeight\gets$ $cellHeight + log(subCellSize) * c $ \Comment{c is a constant determined by the screen size}
\EndFor
\EndFor
\State \Return $cells$
\EndProcedure
\end{algorithmic}
\end{algorithm}

\textbf{Design alternatives.} In our initial design, we chose to build a parallel coordinate diagram for relationship patterns between fields and embedded it in the tabular view. However, the edges caused severe visual clusters. We then switched to a chord diagram, but later some users suggested that the heatmap is more intuitive and helpful in revealing patterns. In addition, there were several alternatives for the tabular structure view, such as sankey diagrams or storyline-based text variant graphs~\cite{silvia2016visualizing}. It turned out that the sankey diagram excels in visualizing a \textit{network} structure within records and fields, while the tabular layout has its advantages in easy navigation and manipulation of the data. We eventually decided to combine these two options by augmenting the tabular layout with edges between cells to facilitate graphical navigation. Users found the result layout to be intuitive and insightful, enabling them to perform data analysis and transformation tasks more efficiently.



\subsubsection{Wrangler view}
The wrangler view is designed for users to construct structured tables from the raw dataset for further tasks (DG5). Columns can be added by dragging cells from the tabular view into the table. Users can freely edit the table contents as needed. Additionally, they can click on individual cells (\autoref{fig:system} (C1)) to locate the corresponding records in the data view (DG3).

\subsection{Interactive wrangling}
 In addition to the visual profiles, users can directly interact on the profiling results for data transformations (DG4). This feature allows users to experiment with the data immediately after gaining insights to promote a seamless workflow. Specifically, to cover the wrangling operations in Section \ref{wrangleop}, StructVizor supports three classes of interactions: visualization-based filter, cell configuration menu, and drag-and-drop.

\textbf{Visualization-based filter.} StructVizor leverages bar charts and histograms to profile the value distributions of fields and subfields (\autoref{fig:system} (A3)). These visualizations enable users to apply filters directly based on the displayed data. By clicking on the chart within cells, users can select specific values to be filtered. The filtered records will be highlighted in both the data view and the histograms above the columns in the structure view as previews. Users can then click the ``\textit{Apply}'' button to apply the filter (\autoref{fig:system} (A5)).

\textbf{Cell configuration menu.} Users can right-click on a cell in the structure view and access the consequent menu (\autoref{fig:system} (A6)) to perform various operations. Informed by DG4 and the transfomation operations in Section \ref{wrangleop}, StructVizor supports the following operators:

\begin{itemize}
    \item \textit{Add.} Select one or more cells to add custom fields or subfields before or after them. The new content can be specified using fixed strings or derived from the selected cell values. Users can also append empty records around the selected cells.

    \item  \textit{Delete.} Delete selected cells, the records associated with them, or entire record clusters.

    \item \textit{Replace.} Transform the values of selected cells by applying operations such as substring extraction, replacement, and regular expression matching.

    \item \textit{Separate.} Split a cell into two by selecting values to place in a new cell, causing the corresponding cells in other columns to reposition accordingly.

    \item \textit{Combine.} Aggregate values of selected cells, replacing the original values or placing the results in a new record.

    \item \textit{Schema.} Modify the schema of subfield cells by editing the regular-expression-based pattern.
\end{itemize}


\textbf{Drag-and-drop.} Drag-and-drop interactions provide an intuitive way to perform operations that involve reordering or repositioning items. For instance, users can easily reposition the records, fields, or subfields by simply dragging and dropping them. Furthermore, StructVizor allows users to construct relational tables in the wrangler view (DG5). By dragging a cell from the structure view to the table in the wrangler view, users can append the values of the corresponding field as a new column. If multiple columns are present in the table, they will automatically be joined based on the raw dataset.

In addition, one interesting usage of drag-and-drop is to decompose complex sub-structures. In practice, data records often include fields with sub-structures, such as  comma-separated items, where multiple instances are consolidated within a single record and thus impede analysis. In such cases, users can drag the cell of one instance to the bottom edge of another, thereby mapping the original record into multiple records, with each instance replacing the field value in a separate record. A concrete example has been given in \autoref{fig:scenario} (C), where the author list are mapped to multiple individual records.
This operation resembles the ``\textit{fold}'' operation commonly used in tabular data wrangling and enables users to effectively disentangle and analyze sub-structures within the data. Note that after each drag-and-drop interaction, a preview dialog window will pop up, displaying the changes from the current transformation (with red text indicating deletions and green text indicating additions).


Besides, it is worth noting that after applying a wrangling operation to the raw dataset, both the data view and the structure view will be updated to ensure the consistency between the profiling results and data (DG3). StructVizor also allows users to undo or redo operations by clicking on the buttons at the interface header.

\subsection{Implementation}
StructVizor was developed following the client-server framework. The frontend interface was developed as a web application using JavaScript and the Vue framework. Additionally, it integrated a backend responsible for processing the input dataset for profiles. The backend was implemented in Python, leveraging Flask for the web server and several third-party libraries such as Scikit-learn, PROSE, and Regex, along with GPT-4o for data manipulation. Specifically, GPT-4o is used for data sampling and parsing, while the libraries are used for structural pattern analysis.

\section{Technical evaluation}


To evaluate the performance of StructVizor's data preprocessing pipeline, we conducted a technical assessment using a substantial set of semi-structured datasets. To the best of our knowledge, we haven't found any open-source benchmark datasets for parsing general semi-structured data except  PADS\footnote{http://www.padsproj.org/learning.html}, which was used in previous studies~\cite{fully-automated, structure-interpretation} but contains only 15 files, mostly log files. To thoroughly test StructVizor's performance across diverse data types, we followed the methodology of UnRavel~\cite{structure-interpretation} and created a new benchmark by gathering 100 open-source semi-structured textual files\footnote{https://github.com/Amur-N/Semi-structured-Dataset-Collection}. Of these, 85 were sourced from GitHub using keywords like \textit{``DB''}, \textit{``txt''}, and \textit{``xml''}, while the remaining 15 files were drawn from the PADS dataset. The file sizes varied from 5KB to 312KB, with 67 files consisting of ad-hoc, non-standardized semi-structured textual data where records can span one or multiple lines. The other 33 files included common semi-structured formats such as JSON, CSV, and XML, along with specialized formats like HMM and FASTQ frequently used in the medical field.
For each dataset, we conducted quintuplicate tests to measure the execution time of the preprocessing algorithm, ensuring that the sample data extracted by GPT-4o was identical for each test. All tests were performed on a 12th Generation Intel(R) Core(TM) i5-1235U processor with a base clock speed of 1.30 GHz and 16 GB of RAM. After completing the tests on all datasets, we quantified the average runtime of the preprocessing algorithms and analyzed their correlation with factors such as dataset size, the number of distinct patterns, and the average number of parsed fields per record after preprocessing. In the following discussion we will highlight our main findings, and we refer readers to the Appendix for detailed analysis results.

\begin{figure*}[t]
\centering
\includegraphics[width=\linewidth]{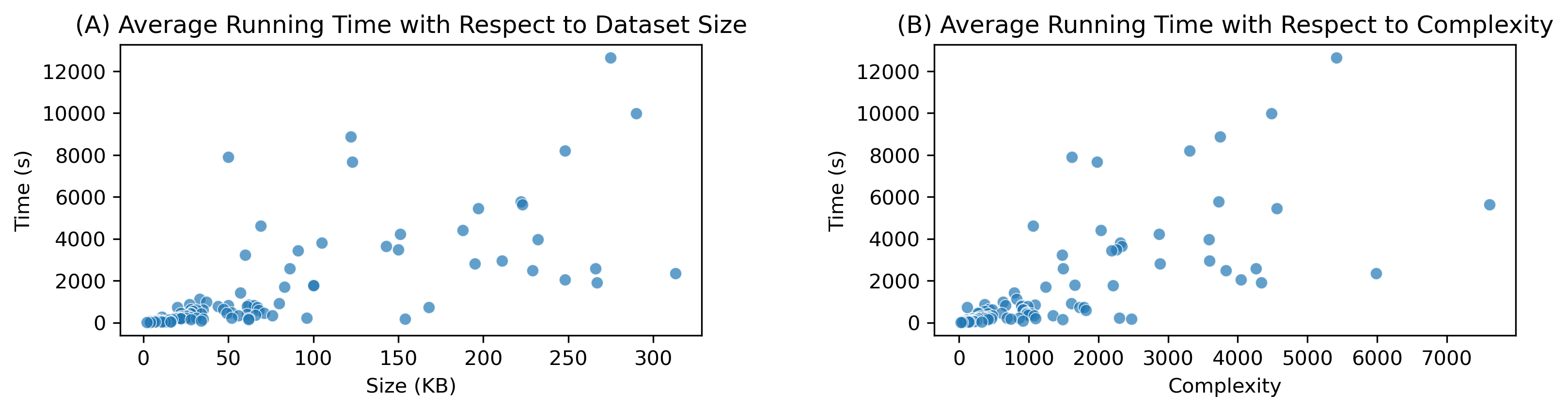}
\caption{Average running time of StructVizor's data processing pipeline with respect to the dataset size and complexity.}
\Description{Overall, the average running time of StructVizor's data processing pipeline grows as the dataset size and complexity increases.}
\label{fig:image1}
\end{figure*}

\subsection{Performance}
\subsubsection{Time} 
\autoref{fig:image1} (A) shows StructVizor's average running time with respect to the dataset size. Overall, the running time is 
higher than the \textit{reported}\footnote{A more formal comparison is infeasible since the tools are not fully open-sourced.} performance of UnRavel~\cite{structure-interpretation} and DataMaran~\cite{datamaran}, which process datasets with several structural assumptions. We argue that this is reasonable as StructVizor's processing pipeline is not restricted by structural assumptions, thus requiring a larger search space. On the other hand, \autoref{fig:image1} (B) illustrates the relationship between the running time and the structural complexity of the dataset, which is characterized by multiplying the normalized value of the the average record length, the number of fields with different structures, and the dataset size. We therefore conclude that the structural complexity of a dataset may significantly influence the processing time. 


\subsubsection{Accuracy}
Additionally, we evaluated the accuracy of the aligned table. Since a dataset can be parsed in a number of ways, an absolute ``ground-truth'' does not exist. Therefore, following prior works~\cite{structure-interpretation}, we categorized all datasets by format into \textit{Log-like files}, \textit{Fixed-width files}, \textit{Key-value files}, \textit{CSV-like files} and \textit{Miscellaneous files}. We selected five datasets from each category for manual verification of the accuracy of the parsed records and fields. Overall, StructVizor successfully parsed most records for \textit{Log-like files} (93.45\%), \textit{Fixed-width files} (81.48\%), \textit{Key-value files} (79.82\%), and \textit{CSV-like files} (94.97\%). The overall accuracy for \textit{Miscellaneous files} is 53.63\%, in which the structural patterns are more random and complicated. By analyzing the errors, we identified two primary factors contributing to these issues: the use of the GPT-4o model and the quality of the sample dataset.

\subsection{Discussion on failures cases}
\subsubsection{Failures from GPT-4o} We noted several errors stemming from GPT-4o's difficulty in accurately identifying record and field boundaries.

\begin{figure}[htbp]
\centering
\includegraphics[width=\linewidth]{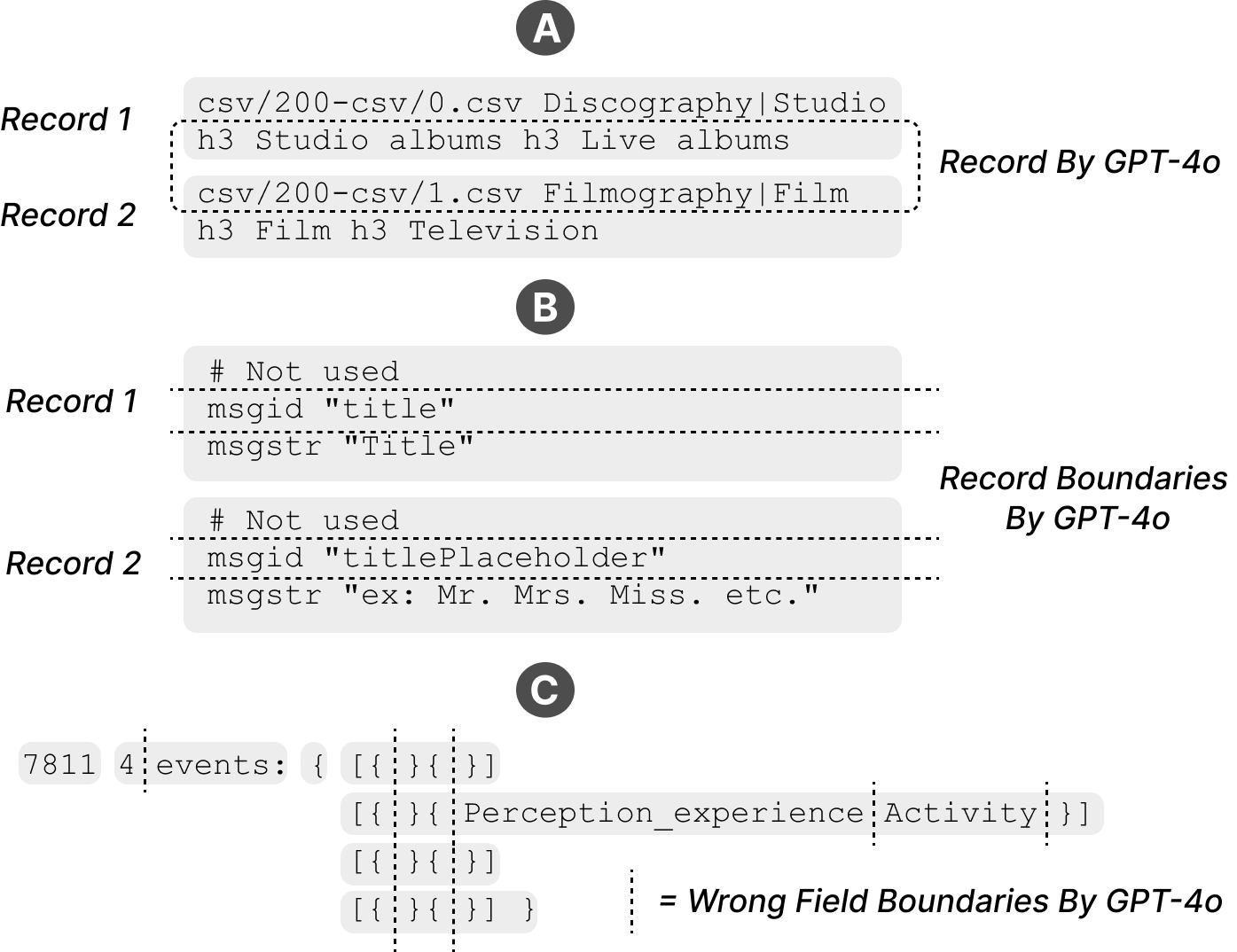}
\caption{Failure cases of GPT-4o in data parsing.}
\Description{In failure case (A), the record parsed by GPT-4o has some positional shift from the ground truth. In failure case (B), the GPT-4o misinterprets the boundaries between records and takes each line as a record. In failure case (C), GPT misinterprets the boundaries between fields and splits the records by the whitespace characters.}
\label{fig:failure}
\end{figure}

(a) Record boundaries: Distinguishing between records requires consideration of both semantic differences and repetitive patterns in adjacent records. GPT-4o may struggle when a clear semantic boundary is absent. For example, in the file shown in ~\autoref{fig:failure} (A), unlike typical files that start each record with an index or ID, this file lacks strong indicators of record beginnings, making each field appear as if it could be the start of a new record. Consequently, records parsed by GPT-4o may have a positional shift from the ground truth.  Additionally, when parsing a section of the file, GPT-4o lacks awareness of the broader context, and the specific part that could provide this context is unknown, leading to these failures. Furthermore, GPT-4o may become overly focused on repetitive patterns, leading to oversplitting of records, as illustrated in ~\autoref{fig:failure} (B). The presence of similar key-value patterns in each record may cause GPT-4o to treat each line as a separate record.

(b) Field boundaries: Contrary to record parsing that requires a broader context, field identification mainly depends on subtle local semantic differences, and parsing failures typically occur in files where fields contain lengthy natural language content or consist of many symbolic strings. In these scenarios, GPT-4o may struggle to decide between syntactic and semantic boundaries, resulting in incorrect splits. For instance, in ~\autoref{fig:failure} (C), the presence of numerous brackets complicates the identification of semantic boundaries, causing GPT-4o to split the record by spaces. This leads to semantic fields being fragmented and paired brackets being separated into different fields.

\subsubsection{Failures from the sample dataset} We have also observed that the sample quality may influence the accuracy of the parsing results. Generally, the more comprehensive and representative the sample is, the fewer unmatched strings will be in the parsing results. Our approach may fall short in files that contain lengthy records or a diverse range of patterns, where including sufficient representative records in the sample becomes difficult. 

\section{User study}

To demonstrate the usability and effectiveness of StructVizor, we  conducted an empirical user study with the primary goal to answer the following research questions:

\begin{itemize}
    \item \textbf{RQ1:} Can users make sense of the data profiles from StructVizor and derive valuable insights?

    \item \textbf{RQ2:} Can users use StructVizor’s interactions to do profile-based data wrangling?

    \item \textbf{RQ3:} How do users perceive and apply the data profiles in explorative data analysis?
\end{itemize}

To address RQ1 and RQ2, we compared StructVizor with Wrangler~\cite{wrangler} on a data wrangling task. We further address RQ1 and RQ3 by asking participants to work on an open-ended explorative task using StructVizor. The study design was approved by the ethics committee of our organization.

\textbf{Justification of baseline selection.} We choose Wrangler as the baseline because it is a representative data wrangling interface that supports diverse data transformations and meanwhile provides some basic profiling features (e.g., bar charts based on data quality metrics). To the best of our knowledge, there are no existing prototypes or product visual interfaces specifically tailored for general semi-structured textual data. Although Wrangler primarily focuses on tabular data, it offers a range of data wrangling operators akin to those in Pandas, which are commonly used by data analysts today. These operators are capable of handling semi-structured textual data to some extent (e.g., one can parse a dataset using functions like \textit{split}, \textit{cut}, and \textit{extract}), representing some of the most prevalent methods for processing such data according to our experience. While modern commercial tools like Trifacta~\cite{trifacta} and OpenRefine~\cite{openrefine} are also valid options, they come with a plethora of detailed features that are hard to master within limited time (especially for first-time users), and they impose specific requirements on the running environment. Therefore, we chose a research prototype as our baseline, leaving a comparison with commercial tools for future work.

\subsection{Study Design}
\subsubsection{Participants} We recruited 12 participants (P1-P12, 7 males and 5 females, aged 22-32) from a local university, including 2 undergraduates, 1 master, and 9 Ph.D. students. Participants reported a high level of experience in data analysis (M=3.83/5, SD=0.83). Three participants indicated they worked with complex semi-structured or unstructured data daily, another three weekly, and six monthly. All participants had experience writing code using open-source libraries such as Pandas. Additionally, seven had used professional systems like Tableau, Trifacta, and SPSS, while nine had experience with AI tools like ChatGPT for analyzing textual data.

\subsubsection{Task} The data wrangling task was based on a semi-structured IMDB dataset consisting of 102 records, with records containing a maximum of 12 fields. We limited the dataset size because Wrangler displays data tables with pagination (25 records per page), making it cumbersome for users to navigate larger tables. 
We chose a file with explicit delimiters (semicolons and whitespaces) to simplify the record and field parsing process. This decision allows us to avoid the complexities of handling \textit{implicit delimiters}~\cite{tegra}, which are poorly supported by existing tools and would have added unnecessary difficulty to our already complex subsequent subtasks. However, parsing using these explicit delimiters may still lead to issues such as insufficient splitting, making the data unready for analysis and requiring users to manually divide certain fields for further processing. Meanwhile, such a file differs from typical structured data, where records are similarly composed and regularly organized, as the parsed dataset included three major classes of data records  with distinct components arranged randomly. Additionally, the records presented several quality issues, such as inconsistent values and noise, which introduced an appropriate level of structural diversity. Participants were instructed to complete a data cleaning task consisting of the following steps: a) remove contextual and problematic records and fields, specifically headers, blank entries, and those containing noise; b) standardize dates of varying formats; and c) split and restructure values from a field containing comma-delimited data  through folding. We ensured these three subtasks were manageable for both systems and encompassed all major categories of text wrangling operations in Section \ref{wrangleop}.

\begin{figure}[t]
  \centering
  \includegraphics[width=\linewidth]{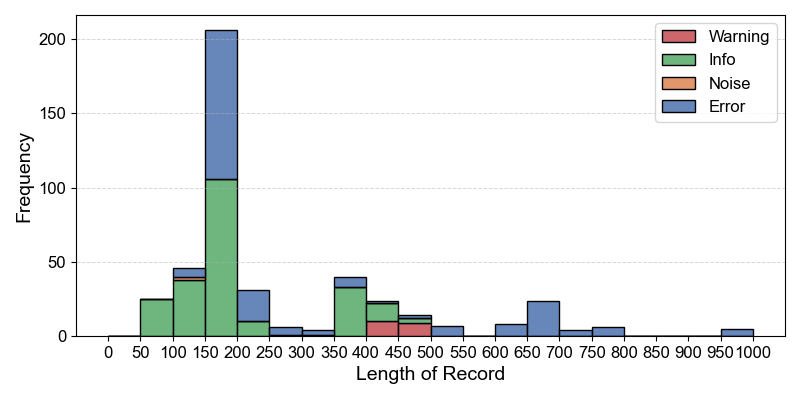}
  \caption{Length distribution of records in the Kubernetes log dataset, categorized by record type.}
  \Description{The dataset include four types of data records. Lengths of records in ``Info'' and ``Error'' type range from 100 to 1000. ``Noise'' records' lengths range from 100 to 150. ``Warning'' records' lengths range from 400 to 500.}
  \label{fig:hist_log}
\end{figure}

In the open-ended explorative task, we used a 127KB log file from Kubernetes, which includes 450 records with their lengths ranging between 63 and 975 characters (approximately 22 first-level fields in the final parsing results of StructVizor).  Records of this file can be broadly classified into four classes: information logs, error logs, warning logs, and noises. However, there are still significant structural differences within each class. \autoref{fig:hist_log} shows the length distribution of the records in each class. Meanwhile, we made sure that there existed interesting positional patterns (e.g., noise records mainly lie in the initial part of the dataset) and field-wise relationship patterns (e.g., recurring strings in different fields), ensuring a diverse pattern space for exploration.

\subsubsection{Procedure} The study started with a demographic survey. Afterwards, participants were asked to finish the task with both StructVizor and Wrangler. To mitigate potential learning effects, we shuffled the order in which participants interacted with the two systems. Before using each system, participants were provided with a 10-minute tutorial, followed by a 5-minute trial period to get themselves familiar with the system. They then worked on the task within a time limit of 15 minutes. Since both systems contained a non-trivial number of functionalities, they were allowed to refer to a manual prepared by us at any time during the study, which included explanations of all detailed interactions of both systems. Upon completion they were asked to finish two questionnaires ($Q_a$ and $Q_b$) assessing their experience with the two systems.
After a 1-minute break, they were given 10 minutes to freely explore the log file using StructVizor, during which they were asked to follow the think-aloud protocol. The study concluded with another questionnaire ($Q_c$) and one follow-up interview. 
The total duration of the study varied between 65 and 80 minutes, and the entire process was video recorded. Participants received a compensation of \$10 after the study.

\subsubsection{Measure} We measured the task completion time for each system during the study. Besides, both questionnaires $Q_a$ and $Q_b$ included six NASA-TLX~\cite{nasa-tlx} questions. To answer the three research questions, $Q_c$ comprised of three questions assessing StructVizor in terms of (a) the comprehensibility of visual profiles, (b) the usability of the interactions, and (c) the helpfulness of profiles for explorative analysis. All questions were measured using 7-point Likert Scales.

\subsection{Quantitative Results}
\subsubsection{Task completion number and time} 
\begin{figure}[t]
  \centering
  \includegraphics[width=\linewidth]{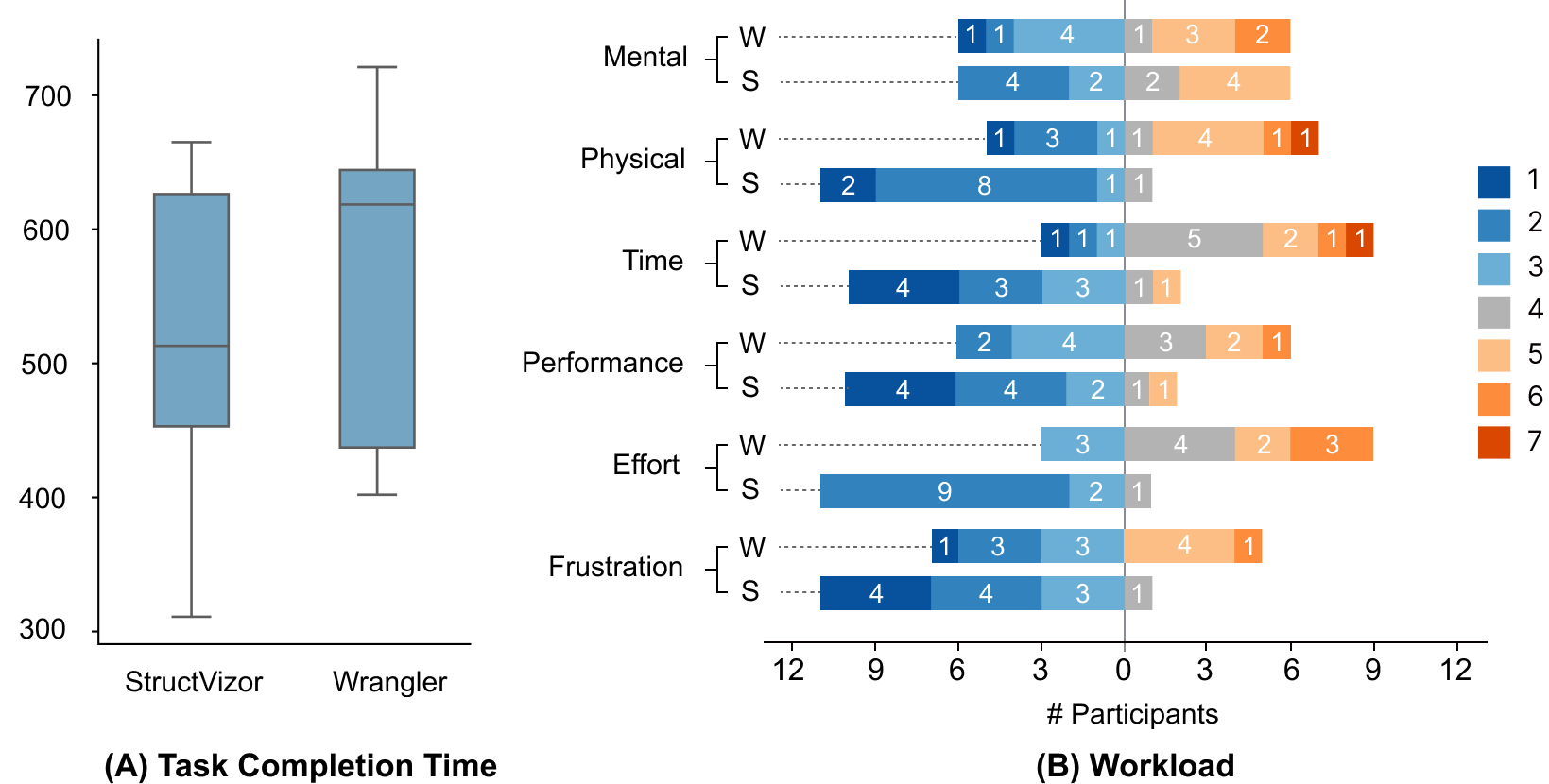}
  \caption{Task completion time (in seconds) and perceived workload of participants for StructVizor (S) and Wrangler (W) on the data wrangling task.}
  \Description{(A) The boxes for the task completion time overlap for both systems, with StructVizor having a lower median. (B) StructVizor's workload is significantly lower than that of Wrangler in all dimensions except mental, where the values are similar. }
  \label{fig:box_time}
\end{figure}

All participants successfully completed the task using both systems within the 15-minute time limit. \autoref{fig:box_time} (A) illustrates participants' time cost. On average, participants took less time when using StructVizor (8 minutes and 36 seconds) than Wrangler (9 minutes and 32 seconds). However, the difference was not statistically significant ($T(11)=-1.461, p = 0.086$, paired student's t-test).

\subsubsection{Workload}
\autoref{fig:box_time} (B) shows participants' self-reported ratings on their perceived workload. Overall, users reported a lower workload across all six NASA-TLX dimensions, with significant differences ($p<0.05$) observed in all dimensions except for \textit{Mental}, which showed no statistical significance ($p=0.60$) in the Wilcoxon signed-rank test. Participants noted in post-study interviews that some found Wrangler's data transformations comparable to those in Pandas and thus more familiar to them, making mental workload lower when using Wrangler compared to StructVizor.

\subsubsection{Effectiveness and usability} In general, participants believed StructVizor's visual profiles easy to understand (M=5.75/7, SD=1.06) and the interactions easy to use (M=6.08/7, SD=0.51). Moreover, all of them agreed that the visual profiles could facilitate their explorative data analysis (M=6.08/7, SD=0.67).

\subsection{Qualitative Feedback} 

\subsubsection{Data profiling} In general, participants were positive about StructVizor's visual profiles. We summarize our major findings as follows.

\textbf{StructVizor promotes data onboarding and analysis planning.}
Most participants (10/12) felt that StructVizor facilitates insights by providing a clear onboarding of the whole dataset, allowing them to identify interesting patterns and fields at an earlier time and better plan their analysis. The tabular overview combined with record clustering was believed to highlight records with varying structures, particularly making those containing issues salient. The heatmap was also recognized as a significant factor, as P10 remarked, ``\textit{Strongly connected fields in the heatmap often indicate redundancy, which helps me filter out less interesting fields.} By contrast, participants felt a lack of table-level profiles apart from column-level ones in Wrangler, leading to five of them feeling ``\textit{overwhelmed}'' (P10) by the number of rows and columns and spending more time locating the target data before starting the given tasks. 

\textbf{StructVizor's expressive profiles facilitates data sensemaking and validation.} All participants believed that  StructVizor's profiles are more expressive than Wrangler, advancing from single data types (e.g., numbers, strings, and null values) to data composition and high-level patterns. For instance, P5 believed that the visual profiles enhanced data validation, sharing an example where Wrangler mistakenly reported the column consistency because it struggled to distinguish between the regular pattern ``\textit{tt\textbackslash d\{7\}}'' of IMDB ID and the noise value ``\textit{abcdefghj}''. However, he could easily check the pattern by double clicking on the corresponding node in StructVizor. P2 and P9 similarly complaint about the frequent unexpected results after transformations and going over the whole table to examine the results in Wrangler.

\textbf{StructVizor improves the efficiency of insight seeking.} Most participants believed that unlike Wrangler, which highlights noteworthy values in a selected column, StructVizor's visual representation of cell blocks merges similar cells of interest (7/12) and encourages them to compare across columns for more insights (3/12). Upon noticing interesting cell blocks or empty sections in the tabular view, users can easily locate the broad context (i.e., records with or without the block) by looking around, helping them finding the reason of issues and efficiently leading to insights.

\textbf{Top-down and bottom-up analysis are complementary and preferred by different users.} Half of the participants (6/12) perceived their data analysis in StructVizor as more systematic, as they can overview the structural patterns through visual profiles, and then navigate to data details in the data view. They felt that such a top-down style of analysis was more ``\textit{natural}'' (P3, P8) and ``\textit{structured}'' (P1). 
However, two participants noted that Wrangler's explicit data table enables them to spot subtle differences that might be overlooked in StructVizor, potentially leading to serendipitous findings. For instance, P6 mentioned that ``\textit{the data table itself is a profile}'', as he browsed the data table in Wrangler and compared adjacent cells by chance for insights. He then moved on to find cells with similar patterns. In StructVizor, getting such detailed insights requires more effort, as users must delve into the hierarchical overview to access field-level or subfield-level information, which is not readily visible at the top layer. 


\textbf{Users expect profiles to seamlessly integrate with data details.} As data analysis is often iterative, users frequently need to navigate between high-level patterns and detailed data. While StructVizor facilitates this process through view coordination, P10 noted that this coordination can be cumbersome due to the frequent need to switch between views. She explained, ``\textit{Switchings between the structure view and the data view happen all the time. I hope they are one view if possible.}'' Furthermore, as one of the most unexpected findings, two participants thought such coordination might cause some cognitive confusion due to the alignment algorithm. P7 illustrated this issue with an example, explaining that the first field of a record, after alignment, might appear in the sixth column of the tabular view instead of the first, contrasting his traditional table-based thinking. 




\subsubsection{Data wrangling} Participants generally expressed a preference for StructVizor's wrangling interactions over those in Wrangler. Five of them attributed this preference to the lower expertise required for StructVizor's interactions compared to Wrangler's data transformation operators. Three participants noted that StructVizor allows for data transformation in the same context where insights are gained, supporting more seamless exploratory analysis.

\textbf{Profile-based interactions are more user-friendly than formal operators.} One of the most frequently mentioned drawbacks of Wrangler was the difficulty in breaking down tasks into a series of operators. These operators were thus perceived as less straightforward than StructVizor's interactions, as the latter resolve tasks in fewer steps. This was particularly evident in the subtask of splitting and folding a comma-delimited field, where many users struggled to organize their thoughts when using Wrangler. We have summarized three reasons.  First, Wrangler requires users to split fields into manageable units, adding an unnecessary ``\textit{preprocessing}'' step (P12), while StructVizor simplifies this with a double-click interaction.
Second, Wrangler's ``\textit{fold/unfold}'' function is a multi-column operator that users find confusing due to ``\textit{abrupt changes}'' (P2) in table structure. In contrast, StructVizor's drag-and-drop approach offers a more intuitive, ``progressive'' (P2) solution.
Third, StructVizor's interactions are based on cell blocks rather than rows or columns, which users found more flexible. This design allows for quick data grouping for different operations right from the start. In Wrangler, users typically apply global changes first and then refine specific problematic values.

\subsubsection{Directions for improvement} Five participants  expressed their desire to see a recommendation view in StructVizor, which, however, was a fundamental feature of Wrangler. They hoped insights could be automatically generated to further enhance the efficiency of the insight-seeking process.  Meanwhile, participants had mixed opinions regarding the learning curve of StructVizor. Four found it easy to learn, while five were neutral; they initially felt overwhelmed by the number of functionalities but later discovered that most were intuitive. Three participants preferred Wrangler, as they found its workflow more natural given their experience with Pandas and spreadsheet tools.  We aim to improve the onboarding and real-time guidance for users in the future.
\section{Discussion}
\subsection{Implications}
\subsubsection{Implications for managing semi-structured textual data.} 
From our design process we learned several important lessons that may implicate semi-structured data analysis. 

\textbf{Boundaries of data type classification.} 
In the data science literature, data has long been classified as \textit{structured data}, \textit{semi-structured data}, and \textit{unstructured data}. However, the definitions of these terms are hardly precise, and it is even more challenging to distinguish between them -- as shown in our paper, structures can be \textit{nested}, and even structured or semi-structured formats may contain a large proportion of unstructured text. Although the ambiguity of these terms' boundaries has been identified as early as 1997 ~\cite{abiteboul1997querying}, few theories have been proposed for clarification since then, particularly due to the difficulty in modeling data that are diverse, heterogeneous, and unpredictable by nature. Recently, researchers have proposed several algorithms to parse or model semi-structured data, though built upon specific structural assumptions~\cite{tegra,datamaran,structure-interpretation}. Our paper extends prior works by proposing a pipeline of data parsing and pattern mining for general semi-structured data, while enriching existing literature by summarizing a space of strctural patterns and wrangling operations. 

\textbf{Profiling semi-structured data.}
Throughout the design process, one of the most significant lessons we learned is that profiles should be tailored based on downstream analysis tasks and specific user needs. For instance, treating an author list field in the citation dataset as a whole works well when reformatting citations, while it is better to dive into the substructure for detailed author analysis. In our approach we propose a hierarchical model of data that include records, fields, and subfields, where the first two levels are precalculated and the others are calculated only when specified by users. 
Overall, such a design ensures the profiles to be dynamic and flexible, successfully helping participants in data sensemaking and fine-grained text wrangling.
However, it remains unclear what kinds of structural patterns useful for different users and tasks, making customized recommendations challenging. For instance, we found that only P10, an analyst who dealed with social media data everyday, checked the string prefix/suffix patterns in her analysis because she often checked them in her daily work. Similarly, few participants checked the area chart in the tabular view, and some explained that this design was irrelevant to their intended analysis tasks but might be potentially useful in other scenarios. We anticipate that large-scale empirical studies of real-world text analysts' practices will yield valuable insights.


\textbf{Measuring data structureness.} Given the ambiguity of data type terms, we envision a quantitative measure of the data structureness as one of the promising future directions. With the help of structural template representations, one can come up with diverse metrics for data consistency and variance. This is especially helpful to give users an initial impression of the difficulty in data analysis, given that in practice many non-professional users often give up their analysis of complex data according to our user study. Additionally, our approach can be further enhanced with \textit{semantic regular expressions}~\cite{semanticregex} for better expressiveness in small- or medium-sized datasets, where semantic metrics will also be helpful.

\subsubsection{Combining data profiles and visual analytics facilitates sensemaking of data.} StructVizor incorporates various visualizations to reveal the mined structural patterns. It extends prior visual profiling interfaces in both objective and dimension~\cite{profiler, datapilot}: it promotes the profiling objective from individual columns (fields) to the entire dataset, and expands the profiling dimension from basic data facts (e.g., quality, type, size) to a more diverse space (e.g., composition, nested structure, data dependency), as introduced in Section \ref{strpattern}. As a result, participants found such visual profiles intuitive and more comprehensive than those in Wrangler.

Among the views, the tabular view is appreciated by most users, as it provides an intuitive overview of data patterns not present in Wrangler, as well as offering a structural approach for data analysis. However, some participants noted that this approach may hinder serendipitous insights through comparing data details. We therefore argue that in visual data analysis, both high-level visual abstractions and concrete instances can be sources of insights and are thus equally essential for users. Additionally, while our data view displays data details, users felt that it could be improved, partly because users prefer an intuitive data table to a plain text view under the influence of spreadsheet or data frame-based tools. This disconnect from traditional workflows may impose an extra mental workload on users, as reflected in a recent study~\cite{TI}. Moreover, we confirm that users need always-on support for both data profiles and details, supporting prior research findings~\cite{dead-or-alive}. While we use view coordinations to connect them, this design might be hard to afford highly frequent switchings between profiles and details, and one potential solution is to merge these views as one view.


\subsubsection{Data profiles as a novel data wrangling paradigm.} \textit{Programming by demonstration} (PBD), which uses command-like operators to execute tasks as adopted in Wrangler, is still one of the most popular data wrangling paradigms. While PBD is formal and unambiguous, its drawbacks include the requirement for users to understand diverse operators that demands high user expertise~\cite{rigel,nl2rigel}, and the need to decompose tasks into long operator sequences~\cite{rigel,foofah}. 

In response, a recent trend, which we term \textit{programming by metaphor} (PBM), has emerged, using intuitive visual metaphors for semantic scope selection and mapping native interactions like drag-and-drops to programming operators, as demonstrated in Table Illustrator~\cite{TI}. Our study follows this trend, demonstrating the opportunities of data profiles on in-situ data wrangling for semi-structured data. Specifically, StructVizor incorporates three classes of interactions in its core tabular view, introducing a new interaction paradigm with cell blocks as the core interaction unit, which are fundamentally a small collection of values with the same context or pattern. This visual approach not only facilitates efficient fine-grained edits, which are prevalent in semi-structured data, but also simplifies complex transformations, such as separation, fine-grained schema modifications, and field switching and folding, through intuitive drag-and-drop interactions. Our user study supports previous findings, highlighting the benefits of combining natural visual hints with visual programming, leading to lower cognitive workload. In the future, we envision a more rigorous formulation of PBM approaches as well as systematic empirical evaluations of them in various domains. Moreover, as recent research use programming by example approaches~\cite{flashprog, foofah} or generative code agents~\cite{copilot} to synthesize code, future work can explore similar support for PBM, recommending visual interaction sequences.


\subsection{Limitations and future work}

First, the current profiling pipeline of StructVizor primarily emphasizes common syntactic patterns and transformations. Future work could explore the integration of external knowledge to identify domain-specific patterns and allow users to customize complex patterns. Additionally, incorporating visualizations for text semantics, such as topic mining and discourse analysis~\cite{interpretation-and-trust, tiara, zhao2012facilitating}, could deepen users' understanding of the data and facilitate more advanced transformations and analyses. 


Second, our current evaluations of StructVizor have some limitations. Since StructVizor is mainly designed for profiling semi-structured data and the interface consists of numerous features, the user study is mainly intended to evaluate the effectiveness of the visual profile design and interactions while touching little on the data processing, and we leave it for future work. While we have technically evaluated data processing, it is hard to systematically evaluate GPT-4o and quantitatively measure the influence of factors such as sample quality and structural complexity on the system performance, due to the large space of data and samples. We believe our approach can benefit from future research on evaluating large language models and better benchmarks of semi-structured data.


Furthermore, StructVizor's data processing pipeline can be improved in many ways. First, as an initial attempt to address the semi-structured data parsing problem without structural assumptions, we envision future work to improve the performance of our algorithms. We suggest that practitioners could divide the entire dataset into smaller batches and leverage parallel computing for more efficient processing. Accordingly, tailored interfaces should be proposed to support interactions for data batches, and we leave this for future work due to the substantial research effort required in the process.  Second, it incorporates a large language model, GPT-4o, in several steps, and the model is well known for its risks such as security and privacy. In StructVizor, the model is only used on the sample set rather than the entire raw data, and we encourage users to remove or replace sensitive data when editing the sample set to avoid privacy or security concerns. Future work may focus on developing more robust privacy-preserving techniques or improving data anonymization methods to ensure user data remains secure. Third, we currently handle potential errors of the data processing pipeline by allowing user edits and rerunning relevant steps, which might lead to significant changes to the results if the user edits are substantial. We plan to improve this approach by deducing  intents from user edits for intelligent refinement of the parsing results.

\section{Conclusion}
We present StructVizor, an interactive visual profiling system for semi-structured textual data. The system incorporates a data processing pipeline that parses input datasets and extracts structural patterns. The profiling results are then visualized within an interactive interface, allowing users to explore patterns and gain insights into the data. StructVizor also leverages the generated profiles to enable interactive in-situ textual data wrangling, facilitating a wide range of data transformations. A user study on 12 participants comparing StructVizor with Wrangler showed that most participants completed the data wrangling tasks more quickly using StructVizor. Additionally, all participants reported significantly lower workload levels with StructVizor except for the mental workload dimension. The study also demonstrated that StructVizor's intuitive data profiles and  interactions for profile-based data wrangling contributed to users' explorative data analysis process. We expect StructVizor's approach to be applied to various scenarios such as log analysis, text reformatting, data cleaning, and social media analytics.

\begin{acks}
  The work was supported
by the National Key Research and Development Program of China (2023YFB3107100) and NSFC (62402421).
\end{acks}

\bibliographystyle{ACM-Reference-Format}
\bibliography{references}










\end{document}